\documentclass[12pt,dvips]{article}

\usepackage{amsmath,amssymb,exscale}
\usepackage{array,multicol}
\usepackage{afterpage,float,flafter}
\usepackage{epsfig,rotating,pifont}
\usepackage{cite}
\@addtoreset{equation}{section} \makeatother

\newcommand{\be}{\begin{equation}}
\newcommand{\ee}{\end{equation}}
\title{
\vspace{.1cm}
\huge{Expanding F-theory}
\vspace{1cm}
\vspace*{.5cm}
\author{
\large \text{Matthew Kleban\footnote{email: mk161@nyu.edu}~~and Michele Redi\footnote{email: redi@physics.nyu.edu}}\vspace{.2cm} \\
\emph{Center for Cosmology and Particle Physics,}\\\emph{Department of Physics, New York University}\\
\emph{4 Washington Place, New York, NY 10003}}}

\date{}
\begin{document}

\maketitle \thispagestyle{empty} \vspace*{.5cm}

\begin{abstract}
We construct a general class of new time dependent solutions of non-linear
$\sigma$-models coupled to gravity. These solutions describe
configurations of expanding or contracting codimension two solitons
which are not subject to a constraint on the total tension. The two
dimensional metric on the space transverse to the defects is
determined by the Liouville equation. This space can be compact or
non-compact, and of any topology. We show that this construction can
be applied naturally in type IIB string theory to find backgrounds
describing a number of 7-branes larger than 24.

\end{abstract}

\newpage
\renewcommand{\thepage}{\arabic{page}}
\setcounter{page}{1}

\section{Introduction}

It is well known that point particles in pure 2+1 dimensional
gravity---or relativistic codimension two objects in higher
dimensional theories---generate a conical space with deficit angle
$\alpha = m$, where $m$ is the mass of the particle (we set $8 \pi
G_N = 1$) \cite{thooft}. If the metric is static, the
two dimensional space transverse to the defects has zero curvature
away from the singularities and is non-compact when the total mass
of the defects is less than $2 \pi$, or compact and of spherical
topology when the total mass is exactly $4 \pi$.

In \cite{inhomo} it was shown that these constraints can be relaxed
by allowing the metric to be time dependent. The space remains
flat outside the defects, but is naturally foliated by two
dimensional locally hyperbolic slices. The constant negative curvature of
this two dimensional space accommodates defects with any total mass,
and allows it to have any topology (subject to some mild restrictions).
Physically these solutions describe 2+1 dimensional cosmologies
uniformly expanding from (or contracting towards) a big bang at
$t=0$.

In this note we will make use of this fact to find expanding
solutions for codimension two $\sigma$-model solitons with tension
and topology not allowed by a static ansatz. This is
possible because the $\sigma$-model solitons (unlike abelian Higgs
model vortices, for example) have a scale invariance which, like
distributions of pressureless point masses, allows 
them to expand uniformly (see also \cite{tolley}).

Our interest in these types of configurations arises from
applications in string theory where $\sigma$-models are ubiquitous.
In particular we will show that this construction allows us to
generalize the stringy cosmic strings of \cite{stringystrings}
to arbitrary numbers of defects and general transverse topology.
This in turn can be exploited to find F-theory \cite{vafa}
backgrounds of type IIB string theory with a number of 7-branes
larger than 24.

Our solutions are reminiscent of those of Refs.
\cite{Townsend:2003fx, ohta, Russo:2003ky, silverstein } (see
\cite{Cornalba:2003kd} for a review and additional references),
which considered time dependent solutions obtained by orbifolding
flat space foliated with negatively curved slices.  In our case the
presence of sources means that the transverse metric does not in
general have constant negative curvature, but in certain limits the
sources are point-like and our solutions are locally flat.  It would
be interesting to investigate the connection with these
works.

\section{Construction}

Our results follow in part from  \cite{inhomo}, which
constructed new gravitational solutions corresponding to an
arbitrary number of point-like codimension two objects coupled to
gravity. To see how this works, recall that a codimension two defect
which is boost invariant along its world volume directions---such as
a straight relativistic cosmic string---generates a locally flat
metric with a conical singularity at the location of the defect.
Choosing a metric ansatz
\begin{equation}
ds^2=-dt^2+dx_{i}dx^{i} + e^{\phi(z,\bar{z})}dz d\bar{z}
\label{ansatz1}
\end{equation}
appropriate for a distribution of parallel defects, the vacuum
Einstein's equations reduce to the Poisson's equation for $\phi$,
with $\delta$-function sources:
\begin{equation}
\partial_z\partial_{\bar{z}}\, \phi= \sum_{i=1}^N
m_i\, \delta^2(z-z_i). \label{poissondelta}
\end{equation}
Since the curvature $R_2$ of the two dimensional space $\Sigma$
parametrized by $(z,\bar z)$ is related to $\phi$ by
$\sqrt{\gamma}R_2(\gamma) = -2i \partial_z\partial_{\bar{z}} \phi$,
eq. (\ref{poissondelta}) implies $R_2=0$ away from the defects. We
can apply the Gauss-Bonnet theorem to $\Sigma$:
\begin{equation}
\int_{\Sigma} R_2 + 2 \int_{\partial \Sigma} K_1 =  4 \pi \chi,
\label{mm6}
\end{equation}
where $K_1$ is the extrinsic curvature of the boundary and $\chi$ is
the Euler character. Using eqs. (\ref{poissondelta}) and (\ref{mm6})
one finds
\begin{equation}
2 \int_{\partial \Sigma} K_1 =  4 \pi \chi - 2 \sum_{i=1}^N m_i.
\label{mm7}
\end{equation}
Since a surface of genus $g$ has $\chi=2-2g$, if the space is
compact and without boundary the only solution has spherical
topology and $\sum m_i = 4 \pi$. Non-compact asymptotically conical
solutions exist when $\sum m_i \leq 2 \pi$.

In \cite{inhomo} it was shown that these restrictions on the deficit
angles can be avoided if the two dimensional metric is uniformly
expanding:
\begin{equation}
ds^2=-dt^2 +dx_{i}dx^{i} + t^2 e^{\phi(z,\bar{z})}dz d\bar{z}.
\label{ansatz2}
\end{equation}
The vacuum Einstein's equations become a
Liouville equation for the conformal factor:
\begin{equation}
\partial_z\partial_{\bar{z}}\, \phi=\frac 1 2\,  e^\phi - \sum_{i=1}^N
m_i\, \delta^2(z-z_i). \label{liouvilledelta}
\end{equation}
This differs from the Poisson equation (\ref{poissondelta}) by the
term $ \frac 1 2 e^{\phi}$, which arises due to the $t^2$
dependence. In the absence of sources the solution of this
equation, $\phi = 2 \ln{\left( 2i  \over z - \bar{z} \right)  }$, is
a metric for the hyperbolic plane $H_2$ with constant negative
curvature. The full metric is flat space written in a two
dimensional hyperbolic slicing. In the presence of defects the space
away from the singularities is still locally flat, but the global
solution is non-trivial.  

The constraint on the total mass is modified because the curvature
term on the left-hand side of (\ref{mm6}) now makes a negative
definite contribution. Integrating the Liouville equation (\ref{liouvilledelta}) and using
(\ref{mm6}) one finds that the volume of
$\Sigma$ is given by
\begin{equation}
V=\sum_{i=1}^N m_i+4\pi (g-1) \label{volumedelta}
\end{equation}
(from here on we will drop the boundary term for simplicity).  Positivity of the volume implies that there are
compact solutions with spherical topology only if $\sum m_i
> 4\pi$, and (at least for closed and compact spaces) in general $V>0$ is the only constraint on the existence of solutions for all topologies (see theorem A of
\cite{troyanov}).

\subsection{Sigma Model Solitons}
\label{gc}

Consider the following action
\begin{equation}
S=-\int \sqrt{-g} \,d^dx \left(\frac 1 2 R +
K_{\tau_i\bar{\tau_j}}\,
\partial_\mu \tau_i
\partial^\mu \bar{\tau}_j\right)
\label{sigmamodel}
\end{equation}
describing a complex non-linear $\sigma$-model coupled to gravity.
Here the fields $\tau_i$ are complex scalars which could arise as
moduli of a compactification from higher dimension, and
$K_{\tau_i\bar{\tau_j}} \equiv
\partial_{\tau_i} \partial_{\bar{\tau}_j} K$ determines the metric
of the target space complex manifold. In $d=4$ this is the bosonic
part of a supersymmetric action.

Under some topological conditions actions of the form (\ref{sigmamodel})  admit soliton solutions
describing codimension two objects \cite{Comtet:1987wi}. To see this one can choose the
metric ansatz (\ref{ansatz2}), and assume the scalars depend only on
the transverse coordinates: $\tau_i = \tau_i(z,\bar z)$. The two
dimensional conformal factor $t^2 \phi(z, \bar{z})$ is time
dependent, but nonetheless it cancels  in the equations of motion
for the scalars:
\begin{equation}
K_{\tau_i \bar{\tau}_k \tau_l}  \left(\partial_z \tau_i \partial_{\bar z}\tau_l+\partial_{\bar z} \tau_i \partial_{z}\tau_l \right)
+ 2 K_{\tau_i\bar{\tau}_k} \partial_z \partial_{\bar z} \tau_i=0.
\end{equation}
These equations are trivially solved by any holomorphic (or
anti-holomorphic) functions $\tau_i(z)$, but one should make sure
that the solution is well defined on the entire manifold spanned by
the scalars. The solutions so obtained correspond to a holomorphic
mapping of the spacetime surface $\Sigma$ into the target space manifold. The energy can be
expressed in a Bogomol'nyi-like form as the integral of the K\"ahler
(1,1) form:
\begin{equation}
E= -\frac {i} 2 \int d^2z \,  \partial_z
\partial_{\bar{z}}K(\tau_i(z),\bar{\tau_j}(\bar{z}))=-\frac{1} 2 \int d^2 \tau_1\dots d^2 \tau_n
\sqrt{\det(K_{\tau_i\bar{\tau}_j})}. \label{energy}
\end{equation}
The right-hand side is essentially the  volume of the target space
times an integer $N$ counting the degree of the mappings
$\tau_i(z)$.  The energy $E$ is positive if the K\"ahler metric is
positive definite (as required by supersymmetry), and with our
normalization corresponds to the total deficit angle.

Having solved the scalar equations of motion, we need to show that
Einstein's equations can be solved consistently. With the time
dependent ansatz (\ref{ansatz2}) they reduce to a single equation:
\begin{equation}
\partial_z\partial_{\bar{z}} \phi= \frac 1 2 e^{\phi}-\partial_z
\partial_{\bar{z}}
K(\tau_i(z),\bar{\tau}_j(\bar{z})), \label{liouville}
\end{equation}
which is similar to (\ref{liouvilledelta}) with the
$\delta$-function sources replaced by the smooth energy
density $\partial_z
\partial_{\bar{z}} K$.

As in the case of point particles, some mild restrictions on the
solutions of eq. (\ref{liouville}) arise depending on the total
energy of the source.
Integrating the Liouville equation and using the definition of $E$ we obtain
\begin{equation}
V=E+4\pi (g-1) \label{volume}
\end{equation}
({\it cf.} eq. (\ref{volumedelta})).  In the static ansatz (\ref{ansatz1}) the left-hand side of eq.
(\ref{volume}) would be zero, and the only allowed compact manifold would be
spherical with energy $E=4 \pi$. The expanding ansatz allows much
more general solutions. For spherical topology the total energy
should again be greater than $4\pi$, and for higher genus surfaces
there is evidently no constraint from Gauss-Bonnet (at least if $E>0$).

Physically we expect solutions of (\ref{liouville}) to exist so long as the right-hand side of (\ref{volume}) is positive. 
As we mentioned previously, existence has been proven when the sources are $\delta$-functions \cite{troyanov}.  
In the smooth case the results of \cite{kazdan}  prove existence  for
arbitrary topology of the transverse space $\Sigma$, at least when
it is closed, compact, $\chi < 0$, and the energy density $\partial_z
\partial_{\bar{z}} K > 0$. This last is guaranteed if the K\"ahler
metric on the target space is positive.  The situation is more complex when $\chi \geq
0$.

\subsection{$CP_1$}
\label{cp1}

As an explicit example, we can apply the construction outlined above
to a  $\sigma$-model with $CP_1$ target space \cite{Cremmer:1978bh}.
The K\"ahler potential is
\begin{equation}
K=a^2 \log[1+\tau\bar{\tau}],
\end{equation}
so that the metric is that of a round sphere. The topological
solitons of this model correspond to mappings of the spacetime
sphere into the target space sphere. The simplest example of such a
map is simply $\tau=z$. From direct integration of (\ref{energy}) we
see that the energy of this configuration is $2\pi a^2$. The general
$N$-vortex solution is given in terms of a rational function,
\begin{equation}
\tau(z)=\frac {P(z)} {Q(z)} \label{rational}
\end{equation}
where $P(z)$ and $Q(z)$ are polynomials of degree $p$ and $q$
without common factors. The energy (\ref{energy}) of these solutions
can be easily evaluated by noting that the mapping (\ref{rational})
covers the target space sphere $N$=max$(p,q)$ times, so that
\begin{equation}
E=2\pi a^2 N.
\end{equation}

An explicit solution of the Liouville equation can be found when
$N=1$. Since this is a trivial mapping from sphere to sphere, the
natural guess is
\begin{equation}
e^{\phi}= \kappa \frac 1 {(1+z\bar{z})^2}. \label{solutioncp1}
\end{equation}
By plugging into eq. (\ref{liouville}) one finds that this is a
solution if
\begin{equation}
\kappa=2 a^2-4.
\end{equation}
From the fact that the metric must be positive it follows that
$a^2>2$, which is nothing but the constraint implied by the
positivity of the volume in eq. (\ref{volume}).

\section{F-theory Revisited}
\label{ftheory}

Arguably the most interesting application of our time dependent
solutions is in the context of string theory. The simplest
case is type IIB string theory, where the relevant $\sigma$-model
is provided by the axion-dilaton, which spans an $\rm
SL(2,\mathbb{R})/U(1)$ manifold. Static codimension two
configurations with varying axion-dilaton are the starting point for
F-theory \cite{vafa}.

At the supergravity level the relevant part of the action is
\begin{equation}
S=-\int \sqrt{-g} \,d^{10}x \left(\frac 1 2 R - \, \frac
{\partial_\mu \tau
\partial^\mu \bar{\tau}}{(\tau-\bar{\tau})^2}\right),
\label{axiondilaton}
\end{equation}
which has the form of (\ref{sigmamodel}) with
$K=-\log[-i(\tau-\bar{\tau})]$. Topological solitons of this
theory corresponding to mappings of the Riemann sphere into the
target space manifold were considered in \cite{stringystrings}. To
construct solutions with finite energy one needs to exploit the
symmetries of the theory by allowing $\tau$ to undergo non trivial $\rm
SL(2,\mathbb{Z})$ monodromies as it varies on the compactification
manifold.\footnote{At the classical level the action
(\ref{axiondilaton}) is $\rm SL(2,\mathbb{R})$ invariant, but
quantum effects break this to the discrete subgroup  $\rm
SL(2,\mathbb{Z})$. This is manifest in the twelve dimensional
picture, where the symmetry is the modular transformation of the
complex structure $\tau$ on the torus.}
Since $\tau$ can be interpreted as
the modular parameter of a ``hidden torus'' associated with a 12
dimensional interpretation of the theory, the solutions so
constructed describe an elliptically fibered manifold obtained by
erecting a torus at each point on the base manifold \cite{vafa}.

To construct these configurations more explicitly it is convenient to
parametrize the torus as the complex surface
\begin{equation}
y^2=x^3+f x +g \label{torus}
\end{equation}
embedded in $\mathbb{C}^2$. The complex numbers
$f$ and $g$ determine the complex structure of the torus:
\begin{equation}
j(\tau)=\frac {4(24 f)^3}{27 g^2+4 f^3}, \label{jfunction}
\end{equation}
where $j(\tau)$ is the modular function mapping the fundamental domain
of  $\rm SL(2,\mathbb{Z})$ to the Riemann sphere. Multi-vortex
solutions are obtained by taking $f$ and $g$ to be polynomials
in the coordinate $z$ of the base manifold. The zeros of the polynomial
\begin{equation}
\Delta=27 g(z)^2+4 f(z)^3
\label{polynomial}
\end{equation}
determine points where the torus
degenerates, but where the space is nonetheless regular
\cite{stringystrings}. Noting that the area of the fundamental
domain of $\rm SL(2,\mathbb{Z})$ is $\pi/6$ in our conventions, it
follows immediately from eq. (\ref{energy}) that the total energy is
\begin{equation}
E=N \frac { \pi} {6}, \label{d7}
\end{equation}
where $N$ is the degree of $\Delta$.

In the type IIB context these configurations
carry $(p,q)$ 7-brane charge \cite{vafa, Gaberdiel:1997ud}.
For the case relevant to the simplest F-theory compactifications the
base of the elliptic fibration is the Riemann sphere, therefore from
eqs. (\ref{d7}) and (\ref{volume}) it follows that it takes
precisely 24 7-branes to close the space. The most general solution
of this type is obtained by taking $f(z)$ and $g(z)$ to be
polynomials of degree 8 and 12 respectively. The manifold so
constructed is an elliptic fibration of $K_3$, and the metric on the
base manifold can be found explicitly by solving the Poisson
equation sourced by the $\sigma-$model \cite{stringystrings}.

When the number of 7-branes is greater than 24 they cannot be
placed on a sphere without some source of negative curvature.
Using the time dependent ansatz (\ref{ansatz2}), however, the
problem is of the form we discussed in the previous section. The
solution for the scalars is the same as eq. (\ref{jfunction}), while
the metric is now determined by the associated Liouville equation
(\ref{liouville}).

The Gauss-Bonnet theorem would require at least $N$=25 defects to
find a compact solution. However in this case there will be branch
cut singularities in $\tau(z)$ \cite{stringystrings, bergshoeff}. For this reason we will focus on the
case $N=6n$, as in these cases the solution is regular.
As we show below, in a certain limit this reduces to the Liouville
equation with point-like singularities.

The number of parameters necessary to specify our expanding
F-theory solutions is similar to the dimension of
the moduli space of static $7$-branes. For the case where the
number of singularities is $6n$ the polynomials $f$ and $g$ have
degree $2n$ and and $3n$. This corresponds to $5n+2$ complex free
parameters. Taking into account conformal transformations and the
fact that the solution only depends on the ratio $f^3/g^2$ we
conclude that the general solution with $6n$ singularities depends
on $5n-2$ complex parameters. Note however that, contrary to the
F-theory case, the volume of the internal space (at given $t$) is
now fixed by eq. (\ref{volume}).

\subsection{Sen's Limit}

A general concern about solutions with varying axion-dilaton is that
the classical configuration might be strongly corrected. This seems
a particularly acute worry for our time dependent configurations
because (as we will discuss) supersymmetry is broken. Since the imaginary part
of $\tau$ controls the string coupling, it follows from eq.
(\ref{jfunction}) that there are regions of space where the string
coupling is large and perturbation theory cannot be applied even in
principle.  While the topological origin of the solutions strongly
hints that they will survive in the full theory, it is useful to
find a limit where these solutions are fully under control.

In \cite{sen} Sen considered a special limit of F-theory where
$\tau$ becomes constant and the solution can be understood as an
orientifold of type IIB string theory. We now show that the same
limit can be used in the time dependent case. To obtain a constant
axion-dilaton one must go to a particular limit of the parameter
space of our solutions where $f,g$ in (\ref{jfunction}) are chosen
so that $f^3/g^2 = \alpha$.
The free parameter $\alpha$ controls the string coupling, and can be
chosen so that the coupling is small. The energy $E$ of the solution
in the limit is still determined  by the degree of the polynomial
$27 g^2+4 f^3$.

In this limit the $\sigma$-model solitons collapse to point-like
singularities corresponding to a deficit angle of $\pi$ each \cite{sen}. For
the static F-theory solutions there are four stacks of six $(p,q)$
7-branes each. The manifold is topologically a sphere
whose metric,
\begin{equation}
ds_2^2= R^2~ \prod_{m=1}^4 |z-z_i|^{-1/2}
\end{equation}
is locally flat but with four singularities. Similarly, if the total
number of singularities is $N=6n$ with $n\geq 5$ we can pack them
into stacks of six each. The metric is then determined by the
Liouville equation with $\delta-$sources (\ref{liouvilledelta}).
Around the singularities the metric is just a reparametrization of
flat space with conical deficit. As a consequence only globally it
is possible to distinguish these solutions from the ones in the
usual F-theory. As in \cite{sen} by looking at the transformation of
the coordinates $(x,y)$ of the torus (\ref{torus}) moving around the
singularities one discovers that there exists a non-trivial
$SL(2,\mathbb{Z})$ transformation
\begin{equation}
 \left(\begin{array}{cc} -1
 & 0 \\
0 & -1
\end{array} \right)
\end{equation}
under which $\tau$ is invariant.  

In the perturbative string theory framework such a monodromy is
associated with the presence of an orientifold plane so that each
singularity is described pertubatively by an O7-plane and four 4
7-branes \cite{sen}.
Since our construction allows an arbitrary number of defects and
locally the space is the same as in \cite{sen}, it is natural to
interpret our solutions as an orientifold of string theory with five
or more O7-planes.  In particular this
might allow one to give a perturbative world-sheet description of
these solutions.\footnote{We should note that since some of the
fields transform under the full $\rm SL(2,\mathbb{Z})$ or its double
cover, global obstructions may in fact require $N=24n$.  We thank
Simeon Hellerman for pointing this out to us.}

Another interesting feature of this solutions is that on a surface with constant negative
curvature $2 \pi$ conical defects (known as parabolic singularities)
become possible. Close to such a singularity the metric is given by
\begin{equation}
e^{\phi}\sim \frac 1{z\bar{z} \left(\log z \bar{z}\right)^2}.
\label{parabolic}
\end{equation}
The proper distance from the singularity at $z=0$ to any finite $z$
diverges, but the volume is finite. Evidently moving two $\pi$
singularities together draws that point out into a cusp. The $\rm
SL(2,\mathbb{Z})$ transformation around these points is trivial. It
would be interesting to study the behavior of string theory close to
these singularities and see whether tachyon instabilities appear
along the lines in Ref. \cite{adams}.\footnote{We would like to
thank Allan Adams for discussions on this point.}

\subsection{Other Topologies}
\label{other}

With the static ansatz the only allowed topology is spherical, but
as we have seen these constraints are avoided when the ansatz is
time dependent.  From eq. (\ref{volume}) we see that there is no
topological constraint for $g>1$, while the toroidal topology
simply needs $E>0$.

We can find generalizations of the solutions of
\cite{stringystrings} to cases where the base of the fibration is a
Riemann surface of genus $g$ by considering mappings between the
fundamental region of $SL(2,\mathbb{Z})$ and the Riemann surface. We
can construct such maps by composing the inverse of the $j$-function
with a map from the Riemann surface into the Riemann sphere. In
particular, if $z=h(\zeta)$ is a meromorphic map from a Riemann
surface $\Sigma$ with coordinate $\zeta$ into the complex plane,
then $\tau(\zeta) = j^{-1}\left( h(\zeta) \right)$ describes a
``stringy cosmic string'' configuration with transverse space
$\Sigma$.

This can be done rather explicitly for the case of toroidal
topology. In this case the functions $f$ and $g$ in
(\ref{jfunction}) must be mereomorphic functions defined on the base
torus, i.e. elliptic functions. These are in general rational
functions of the Weierstrass $\wp-$function (with periods determined
by the torus) and its derivative. The simplest solution is given by
\begin{equation}
j(\tau(z))=\frac {4(24 \wp(z))^3}{27 +4 \wp(z)^3}. \label{six}
\end{equation}
Since $\wp$  has one double pole at $z=0$ and two single zeros in
the fundamental region of the spacetime torus (not to be confused
with the ``hidden torus'' of the 12 dimensional description) one can
check that $\tau$ does not have orbifold singularities and is well
defined on the torus. The mapping (\ref{six}) covers the fundamental
region of $\rm SL(2,Z)$  six times (this follows
 from the fact that the $\wp(z)$ is a double cover of the
Riemann sphere), so evidently this configuration can be understood as
six  7-branes on an expanding torus.  We can find an analog
of the Sen limit that should connect these solutions to
perturbative string theory. This could be done for example by taking
$f=\alpha \wp^2$ and  $g=\wp^3$. It would be interesting to study
these configurations more in detail.

We can also find solutions in cases where the transverse
space is non-compact. A simple example follows from the above: given
a solution with toroidal topology, we can always re-interpret the
configuration as an infinite expanding array of defects.   More general configurations
should also exist, either with a finite or infinite number of defects.  
In the case where the total mass is finite, the extrinsic curvature term in eq. (\ref{mm7}) compensates
for the bulk curvature contribution.

\section{Corrections}

F-theory backgrounds constructed from elliptic fibrations of
Calabi-Yau manifolds are supersymmetric \cite{vafa,morrison}. In the
expanding case the time dependence implies that supersymmetry is
broken. It is interesting to see in detail how this happens. For
simplicity we consider the problem in $4D$ with four supercharges
following \cite{becker}. For a supersymmetric $\sigma-$model coupled
to gravity the variations of the fermions are
\begin{eqnarray}
\delta_{\epsilon} \chi_i&=& i \sqrt{2} \sigma^{\mu} \bar{\epsilon}\partial_{\mu} \tau_i \nonumber \\
\delta_{\epsilon} \psi_{\mu}&=& \partial_{\mu} \epsilon- \epsilon
\omega_{\mu}-\frac 1 4 \left(K_{\tau_j}\partial_{\mu}
\tau_j-K_{\bar{\tau_j}}\partial_\mu \bar{\tau_i}\right) \epsilon
\label{variations}
\end{eqnarray}
where $\omega_{\mu}$ is the spin connection. For holomorphic
solutions of the scalars the variation of the spinors $\chi_i$ is
zero for $\hat{\epsilon}=\gamma^{\bar{z}} \epsilon$.
The gravitino equation implies an integrability equation,
\begin{equation}
\left[D_\mu,D_\nu\right]\hat{\epsilon}=0
\end{equation}
where $D_{\mu}$ is defined by the variation of the gravitino above.
On can check that this condition is just the Liouville equation
(\ref{liouville}), so it is automatically satisfied by the
background. However the integrability condition is necessary but not
sufficient to guarantee the existence of Killing spinors. In fact
writing explicitly the variations of the gravitino, due to the time
dependence of the ansatz, one finds that $e^{\phi}\hat{\epsilon}=0$.
This proves that the expanding solutions are not supersymmetric.
However, the energy density dilutes with the expansion and therefore
the supersymmetry is approximately restored at late times.

In Sen's limit,  the background around the singularities is
identical to the supersymmetric case, so locally one can find
solutions of the Killing spinor equations.   There are however no
globally well defined Killing spinors (except possibly in the
non-compact case $N=6$). The breaking of supersymmetry here is
reminiscent of Scherk-Schwarz compactifications where supersymmetry
is broken non-locally by boundary conditions.

Since the $2\pi$ singularities have no flat space analog, we expect
them to break supersymmetry locally and they are likely unstable.
Similarly, configurations of $N$ strings where $N \neq 6n$ are
non-supersymmetric even in flat space, and are likely to be locally
unstable (but see \cite{bergshoeff} for an interesting exception in
the non-compact case).
Since supersymmetry is broken at loop level there will be a force
between the different stacks of 7-branes. At late times this effect
could just be computed from the Casimir energy in the supergravity
approximation. As we approach $t=0$ this will presumably turn in a
tachyon instability similar to the one of the $2\pi$ singularities.

Finally our solutions will have both $g_s$ and $\alpha'$
corrections. As we argued above the first ones become under control
everywhere in the space in the Sen's limit. For the latter it is
sufficient to note that the curvatures scale as,
\begin{equation}
R\propto 1 /{t^2}
\end{equation}
At large $t$ all the correction due to higher powers of the
curvatures become negligible. At least in the Sen's limit, the space
is locally supersymmetric, the string coupling can be taken
arbitrarily small, and the non-supersymmetric states become very
massive, so we expect these solutions to be under good perturbative
control.

\section{Discussion}

In this note we have presented new $\sigma$-model solutions
corresponding to supermassive codimension two solitons coupled to
gravity. The time dependence of the metric allows us to construct
solutions with an arbitrarily large number of defects and differing
transverse topologies, evading the constraints which apply to static
configurations. In particular we have shown that this construction
can be applied to type IIB string theory, generalizing F-theory
backgrounds to an arbitrary number of 7-branes.

We have left many interesting questions unanswered. For example in
F-theory, solutions can be viewed as a compactification from 12
dimensions. In particular elliptically fibered Calabi-Yau manifolds
of any dimension can be used \cite{morrison}. Here we have
considered the simplest case where the base space is two dimensional, and
the Liouville equation plays a key role. It would be interesting to
consider generalizations to higher dimensional base manifolds.
Another direction is to study the field theory description of these 
solutions using brane probes \cite{sen, Banks:1996nj, Sen:1996sk}.

On the gravitational side, one could study the
stability of these solutions under perturbations, for example where
the defects are given some non-zero relative velocities. Perhaps the
most pressing question is what happens as we approach the
singularity at $t=0$. Here higher curvature corrections will become
important and the supergravity approximation breaks down.  Since
supersymmetry is broken instabilities are likely to appear. One
possibility is that tachyon condensation could resolve the
singularity. A related conjecture is that these backgrounds are dual
or connected  via tachyon condensation to supercritical string
theories \cite{Hellerman:2006nx}. We leave these investigations
to future work.

\section{Acknowledgments}

We would like to thank Allan Adams, Ofer Aharony, Eric Bergshoeff,
Jose Blanco-Pillado,   Albion Lawrence, John McGreevy, Massimo
Porrati, Oriol Pujolas, Ra\'ul Rabad\'an,  Augusto Sagnotti,  Eva
Silverstein,and especially Simeon Hellerman for very useful
conversations. M.R. would like to thank J.J. Blanco-Pillado for
collaboration at the initial stages of this work. The work of M. R.
is supported by the NSF grant PHY-0245068.

\vspace{0.5cm}

%%%%%%%%%%%%%%%%%%%%%% Bibliography %%%%%%%%%%%%%%%%%%%%%%%%%%%%%%%%%%%

\end{document}